\RequirePackage{lineno}
\documentclass[showpacs,twocolumn,groupedaddress]{revtex4}
\usepackage{graphicx}% Include figure files
\usepackage{dcolumn}% Align table columns on decimal point
\usepackage{bm}% bold math

\begin{document}

\title{Microinstabilities at perpendicular collisionless shocks:
A comparison of full particle simulations 
with different ion to electron mass ratio}

\author{Takayuki Umeda}
\email[Email:]{umeda@stelab.nagoya-u.ac.jp}
\affiliation{Solar-Terrestrial Environment Laboratory, 
Nagoya University, Nagoya 464-8601, JAPAN}

\author{Yoshitaka Kidani}
\affiliation{Solar-Terrestrial Environment Laboratory, 
Nagoya University, Nagoya 464-8601, JAPAN}

\author{Shuichi Matsukiyo}
\email[Email:]{matsukiy@esst.kyushu-u.ac.jp}
\affiliation{Earth System Science and Technology, 
Kyushu University, Kasuga 816-8580, JAPAN}

\author{Ryo Yamazaki}
\email[Email:]{ryo@phys.aoyama.ac.jp}
\affiliation{Department of Physics and Mathematics, 
Aoyama Gakuin University, Sagamihara 252-5258, JAPAN}

%\newpage

\begin{abstract}
A full particle simulation study is carried out 
for studying microinstabilities generated at 
the shock front of perpendicular collisionless shocks. 
The structure and dynamics of shock waves are 
determined by Alfven Mach number and plasma beta, while 
microinstabilities are controlled by 
the ratio of the upstream bulk velocity to 
the electron thermal velocity and the plasma-to-cyclotron frequency. 
Thus, growth rates of microinstabilities 
are changed by the ion-to-electron mass ratio, even with 
the same Mach number and plasma beta. 
The present two-dimensional simulations show that 
the electron cyclotron drift instability is dominant 
for a lower mass ratio, and 
electrostatic electron cyclotron harmonic waves are excited. 
For a higher mass ratio, 
the modified two-stream instability is dominant 
and oblique electromagnetic whistler waves are excited, 
which can affect the structure and dynamics of collisionless shocks 
by modifying shock magnetic fields. 
\end{abstract}

\pacs{
52.35.Tc; %Shock waves and discontinuities 
52.35.Qz; %Microinstabilities
52.65.-y; %Plasma simulation
52.65.Rr; %Particle-in-cell method 
}

%\received{January 20, 2012}
%\revised{}
%\accepted{}
%\published{}

\maketitle

%\linenumbers*[1]

\section{Introduction}

Collisionless shocks have been investigated by 
full particle (Particle-In-Cell) simulations 
for more than four decades since early 1970's \cite{Biskamp_1972}. 
The full particle method handles 
both electron-scale microphysics and ion-scale shock non-stationarity 
simultaneously 
because both electrons and ions are treated as individual charged particles. 
However, a reduced ion-to-electron mass ratio ($m_i/m_e$) is commonly used 
in the full particle method to reduce the computational cost. 

It is well known that structures and dynamics of shock waves are 
determined by the following two dimensionless parameters, i.e., 
Alfven Mach number 
\begin{equation}
\label{eq:mach}
M_A = \frac{u_{1}}{c}\frac{\omega_{pi1}}{\omega_{ci1}}
    = \frac{u_{1}}{c}\frac{\omega_{pe1}}{\omega_{ce1}}\sqrt{\frac{m_i}{m_e}}.
\end{equation}
and the ratio of the thermal plasma pressure to the magnetic pressure 
(plasma beta)
\begin{equation}
\beta_e = \frac{2v_{te1}^2\omega_{pe1}^2}{c^2\omega_{ce1}^2}, \ \ \ 
\beta_i = \frac{2v_{ti1}^2\omega_{pi1}^2}{c^2\omega_{ci1}^2}, 
\end{equation}
where $c$, $u$, $\omega_{p}$, $\omega_{c}$, $v_{t}$ and $m$
represent the speed of light, upstream bulk velocity, 
plasma frequency, cyclotron frequency, thermal velocity and mass, 
respectively, with the subscripts ``$i$'' and ``$e$'' 
being ion and electron, respectively. 
Here the subscript ``1'' denotes upstream. 

It is well known that 
a part of incoming ions are reflected at the shock front of 
perpendicular and quasi-perpendicular shocks. 
The reflection of ions results in the deceleration of incoming electrons 
so that the conservation of the total current 
(the zero current condition in the shock normal direction) is satisfied. 
Consequently, a relative bulk velocity 
between the incoming electrons and 
the incoming/reflected ions arises. 
Previous self-consistent kinetic simulation studies revealed that
there exist various types of microinstabilities 
at the shock front of 
perpendicular and quasi-perpendicular shocks 
in which reflected ions
play essential roles in dissipative processes. 
The free energy source of these microinstabilities 
is the relative drift (current) between incoming electrons 
and incoming/reflected ions. 

The ratio of the relative drift velocity 
to the electron thermal velocity is important for 
controlling the type of microinstabilities 
(or the effect of electron thermal damping to waves). 
This ratio is proportional to 
the ratio of the upstream bulk velocity 
to the electron thermal velocity. 
From the equations of Alfven Mach number and electron beta, 
we obtain
\begin{equation}
\frac{u_{1}}{v_{te1}} 
= M_A\frac{\sqrt{2}}{\sqrt{\beta_e}}\sqrt{\frac{m_e}{m_i}}.
\end{equation}
This relation means that 
the ratio of the upstream bulk velocity to 
the electron thermal velocity becomes larger 
with larger Alfven Mach number, smaller electron beta, 
or smaller mass ratio. 
Note that the actual amount of free energy 
relative to the electron thermal energy 
($m_i u_{x1}^2/m_e v_{te1}^2$) is 
independent of the mass ratio. 

If the relative velocity between electrons 
and ions exceeds the electron thermal velocity, 
electrostatic waves are excited by current-driven instabilities 
such as the Buneman-type instability (BI) \cite{Buneman_1958} or 
the electron cyclotron drift instability (ECDI) 
\cite{Wong_1970,Forslund_1970}. 
At high-Mach-number perpendicular shocks, 
the relative velocity between incoming electrons and 
reflected ions commonly becomes much faster 
than the electron thermal velocity. 
Then, the BI becomes dominant, and 
electrostatic waves are excited 
at the upper hybrid resonance frequency  
\cite{Shimada_2000}. 
At lower-Mach-number ($M_A < 10$) perpendicular shocks, 
the relative velocity between incoming electrons and 
incoming/reflected ions becomes close to 
the electron thermal velocity. 
Then, the growth rate of the BI 
becomes small because of damping by thermal electrons, 
and the ECDI becomes dominant, which excites 
electrostatic waves at 
multiple electron cyclotron harmonic frequencies 
\cite{Muschietti_2006}. 
When the relative velocity between incoming electrons and 
incoming/reflected ions becomes slower 
than the electron thermal velocity 
at lower-Mach-number perpendicular shocks, 
high-frequency electrostatic waves are not excited 
due to damping by thermal electrons, and 
the modified two-stream instability (MTSI) becomes dominant 
\cite{Lashmore_1971,Krall_1971,Ott_1972,McBride_1972a,McBride_1972b}. 
Then, obliquely propagating electromagnetic whistler mode waves 
are excited at a frequency 
between the electron cyclotron frequency and 
the lower hybrid resonance frequency  
\cite{Wu_1983,Scholer_2003,Matsukiyo_2003,Scholer_2004,Matsukiyo_2006,Matsukiyo_2010,Umeda_2012}.

The previous theoretical work has shown that the growth rate of the MTSI 
is strongly affected by the ion-to-electron mass ratio $(m_i/m_e)$, 
but not by the ratio of the electron plasma to cyclotron frequency 
$(\omega_{pe}/\omega_{ce})$ \cite{Matsukiyo_2003}. 
It has also been confirmed 
by the previous one-dimensional (1D) full particle simulation studies 
that the mass ratio 
affect the shock dynamics as well as the microinstability 
\cite{Scholer_2003,Scholer_2004}. 
In the present study, we vary the mass ratio and frequency ratio 
as shown in Table I. 
Even when the Alfven Mach number and the electron beta are fixed, 
the ratio of the plasma frequency to cyclotron frequency 
($\omega_{p}/\omega_{c}$) 
can take an arbitrary value, because 
$c/v_{te}$ or $u_{1}/c$ can also take arbitrary values. 
The frequency ratio may also control the type of microinstabilities 
by determining the linear dispersion relation of 
background plasma. 
It should be noted that the effect of the frequency ratio 
is not included in hybrid simulations 
where ions are treated as particles while electrons are treated 
as a (massless) fluid. 
That is, the ion inertial length ($c/\omega_{pi}$) and 
the ion gyro radii ($v_{ti}/\omega_{ci}$ and $u_{1}/\omega_{ci}$)
are defined but 
$\omega_{pi}/\omega_{ci} \rightarrow \infty$ and 
$c/v_{ti} \rightarrow \infty$ are assumed in hybrid simulations. 
Thus the full particle simulation is the unique approach 
to study the effect of both mass ratio and frequency ratio. 

In the present study, we perform a series of two-dimensional (2D)
full particle simulations of a perpendicular 
(the shock normal angle $\theta_{B_n} = 90^{\circ}$) 
collisionless shock with a low Mach number ($M_A=6$) 
and a moderate beta ($\beta_i=\beta_e=0.32$).  
We examine the impact of the mass ratio 
on microinstabilities at the shock front 
by fixing either $\omega_{pi1}/\omega_{ci1}$ or $\omega_{pe1}/\omega_{ce1}$ 
as shown in Table I.

\begin{table*}[t]
\caption{
Parameters for different simulation runs. 
}
\begin{tabular}{l|c|c|c|c|c|c|c}
Run \ & $\ m_i/m_e \ $ & 
$\ \omega_{pe1}/\omega_{ce1} \ $ & 
$\ \omega_{pi1}/\omega_{ci1} \ $ & 
$\ u_{x1}/v_{te1} \ $ & $\ c/v_{te1} \ $ &
$\ l_{i1}/\lambda_{De1}\ $ \\ \hline
\ \ A$^a$ & 625 &  4.0  & 100.0 & 0.6    & 10.0   & 250 \\
\ \ B     & 256 &  4.0  &  64.0 & 0.9375 & 10.0   & 160 \\
\ \ C     & 100 &  4.0  &  40.0 & 1.5    & 10.0   & 100 \\
\ \ D     &  25 &  4.0  &  20.0 & 3.0    & 10.0   &  50 \\
\ \ E     & 256 &  6.25 & 100.0 & 0.9375 & 15.625 & 250 \\
\ \ F     & 100 & 10.0  & 100.0 & 1.5    & 25.0   & 250 \\
\ \ G     &  25 & 20.0  & 100.0 & 3.0    & 50.0   & 250 \\
\end{tabular}

$^a$ Umeda \textit{et al.} \cite{Umeda_2012}

\end{table*}

\section{Simulation Setup}

We use a 2D electromagnetic full particle code 
in which the full set of Maxwell's equations and 
the relativistic equation of motion for individual electrons and ions 
are solved in a self-consistent manner. 
The continuity equation for charge is also 
solved to compute the exact current density 
given by the motion of charged particles \cite{Umeda_2003}. 
In the present simulation, 
the simulation domain is taken in the shock-rest frame 
\cite{Umeda_2006,Umeda_2008,Umeda_2009}. 
The shock-rest-frame model is achieved by using the ``relaxation method'' 
\cite{Leroy_1981,Leroy_1982,Muschietti_2006}, 
in which a collisionless shock is excited by an 
interaction between a supersonic plasma flow and 
a subsonic plasma flow moving in the same direction. 

The initial state consists of two uniform regions 
separated by a discontinuity. 
In the upstream region that is taken in the left-hand side 
of the simulation domain, 
electrons and ions are distributed uniformly in space and 
are given random velocities $(v_x,v_y,v_z)$ to approximate 
shifted Maxwellian momentum distributions 
with the drift velocity $\vec{u}_1$, 
number density $n_{1} \equiv \epsilon_0 m_e \omega_{pe1}^2 / e^2$, 
isotropic temperatures $T_{e1} \equiv m_e v_{te1}^2$ and 
$T_{i1} \equiv m_i v_{ti1}^2$, 
where $e$ is the charge. 
Here, the subscripts ``1'' and ``2'' denote 
``upstream'' and ``downstream'', respectively. 
The upstream magnetic field $\vec{B}_{01}$ 
with a magnitude of $m_e \omega_{ce1}/e$ 
is also assumed to be uniform. 
The downstream region taken in the right-hand side 
of the simulation domain is prepared similarly with 
the drift velocity $\vec{u}_2$, density $n_{2}$, 
isotropic temperatures $T_{e2}$ and $T_{i2}$, 
and magnetic field $\vec{B}_{02}$. 

We take the simulation domain in the $x$-$y$ plane 
and assume an in-plane shock magnetic field ($B_{y0}$). 
The shock-normal magnetic field $B_{x0}$ is set to be zero. 
That is, the shock-normal angle $\theta_{B_n}$, 
which is defined as the angle between 
upstream magnetic field and shock-normal direction, 
%= \arctan (B_{y01}/B_{x0})$ 
is set as $90^{\circ}$ in the present study. 
As a motional electric field, a uniform external electric field 
$E_{z0} = u_{x1}B_{y01} = u_{x2}B_{y02}$ 
is applied in both upstream and downstream regions,  
so that both electrons and ions drift along the $x$ axis. 
At the left boundary of the simulation domain in the $x$ direction, 
we inject plasmas with the same quantities 
as those in the upstream region, 
while plasmas with the same quantities as those 
in the downstream region are also injected from the right boundary 
in the $x$ direction. 
We use absorbing boundaries 
to suppress non-physical reflection of electromagnetic waves at 
both ends of the simulation domain in the $x$ direction 
\cite{Umeda_2001}, 
while the periodic boundaries are imposed in the $y$ direction. 

In the relaxation method, 
the initial condition is given by solving 
the shock jump conditions (Rankine-Hugoniot conditions) for 
a magnetized two-fluid isotropic plasma 
consisting of electrons and ions \cite{Hudson_1970}. 
In order to determine a unique initial downstream state, 
we need given upstream quantities 
$u_{x1}$, $\omega_{pe1}$, $\omega_{ce1}$, $v_{te1}$, 
$v_{ti1}$ and $T_{i2}/T_{e2}$.

In the present study, 
we fix the Mach number and plasma beta to 
$M_A=6$ and $\beta_e=\beta_i=0.32$. 
As shown in Table I, we perform 7 different runs 
by varying the mass ratio ($m_i/m_e$)
and frequency ratio. 
We fix $\omega_{pe}/\omega_{ce}$ in Runs B-D. 
Then, the electron parameters are fixed 
while the ion parameters are changed. 
On the other hand, we fix $\omega_{pi}/\omega_{ci}$ in Runs E-G. 
Then, the ion parameters are fixed 
while the electron parameters are changed. 

The size of the simulation domain is 
$L_x = 32l_{i1}$ and $L_y = l_{i1}$ 
in the shock-normal and shock-tangential directions, respectively, 
where $l_{i1} = c/\omega_{pi1}$ 
is the ion inertial length. 
The grid spacing is set to be 
$\Delta x = \Delta y = \lambda_{De1}$ 
where $\lambda_{De1}$ is the electron Debye length upstream. 
On the other hand, the time step is set to be $c\Delta t/\Delta x=0.5$. 
These mean that the computational resource 
(the size of the simulation domain and the number of time step) 
increases with the mass ratio when the electron parameters 
are fixed (Runs A-D), 
while the computational resource becomes 
constant when the ion parameters are fixed (Runs E-G and A). 
We used 16 pairs of electrons and ions per cell in the upstream region 
and 64 pairs of electrons and ions per cell in the downstream region, 
respectively, at the initial state.

\section{Results}

\subsection{Simulation Results}

Figure 1 shows the evolution of the perpendicular shock 
in Runs A-G. 
The snapshots of 
the magnetic field $B_x$, $B_y$, $B_z$ components, 
the electric field $E_x$ component, and the corresponding 
$x-v_x$ phase-space distribution function of ions 
at $\omega_{ci1}t = 10.2$ in Runs A-G 
are plotted in Figures A-G, respectively. 
The position is normalized by 
the ion inertial length $l_{i1} = c/\omega_{pi1}$. 
The magnitudes of magnetic field and electric field are 
normalized by the 
initial upstream magnetic field $B_{y01}$ and 
the motional electric field $E_{z0}$, respectively. 
In the panel (a'), 
the tangential component of the magnetic field $B_y$ in Run A 
is plotted as a function of position $x$ and time $t$. 
Note that the tangential component of the magnetic field $B_y$ 
in Runs B-G shows almost the same evolution as the panel (a'). 

In the present shock-rest-frame model, 
a shock wave is excited by the relaxation of 
the two plasmas with different quantities. 
Since the initial state is given by the shock jump conditions 
for a ``two-fluid'' plasma consisting of electrons and ions, 
the kinetic effect is excluded in the initial state. 
Thus the excited shock slightly moves upstream. 
The shock front appears and disappears on a timescale 
of the ion cyclotron period in Figure 1a', 
which corresponds to the cyclic shock reformation 
\cite{Biskamp_1972,Quest_1985,Lembege_1987}. 

In all the runs, a certain part of the incoming ions 
is reflected upstream at the shock ramp 
and the shock foot region is formed, 
which is called the broadening phase of the reformation. 
Consequently, microinstabilities are generated due to the relative drift 
between the reflected ions and the incoming electrons, and 
ion phase-space vortices in association with wave generation 
are formed as seen in the ion $x-v_x$ phase-space plots. 
The spatial size of the foot region 
corresponds to the gyro radius of reflected ions, 
which seems to be longer with respect to the ion inertial length 
as the mass ratio becomes smaller 
when the electron parameters are fixed (Runs A-D). 
It is found that the wavelength of the magnetic field $B_z$ component 
in the shock normal direction becomes 
longer as the mass ratio becomes smaller. 
The excited waves in $E_x$ and $B_z$ components 
have the same wavelength in Runs A, B, C, E, and F, 
indicating the excitation of electromagnetic waves. 
Also, the wavelength is shorter with the fixed $\omega_{pi}/\omega_{ci}$ 
than with the fixed $\omega_{pe}/\omega_{ce}$.
However, the saturation amplitude of these waves 
is almost independent of the mass ratio 
(typically $\delta B_z^2 \sim 0.3 B_0^2$). 
On the other hand, the excited waves in Runs D and G 
have a shorter wavelength in $E_x$ component 
than in $B_z$ component, 
indicating the excitation of electrostatic waves 
in the electric field $E_x$ component. 
The typical timescale of the saturation of instabilities 
is $\sim 0.6/\omega_{ci1}$ from 
the beginning of the broadening phase (ion reflection). 
Note that the typical timescale of the reformation is 
$\sim 1.6/\omega_{ci1}$ in the present case.

\begin{figure*}[p]
\center
\includegraphics[width=1.0\textwidth]{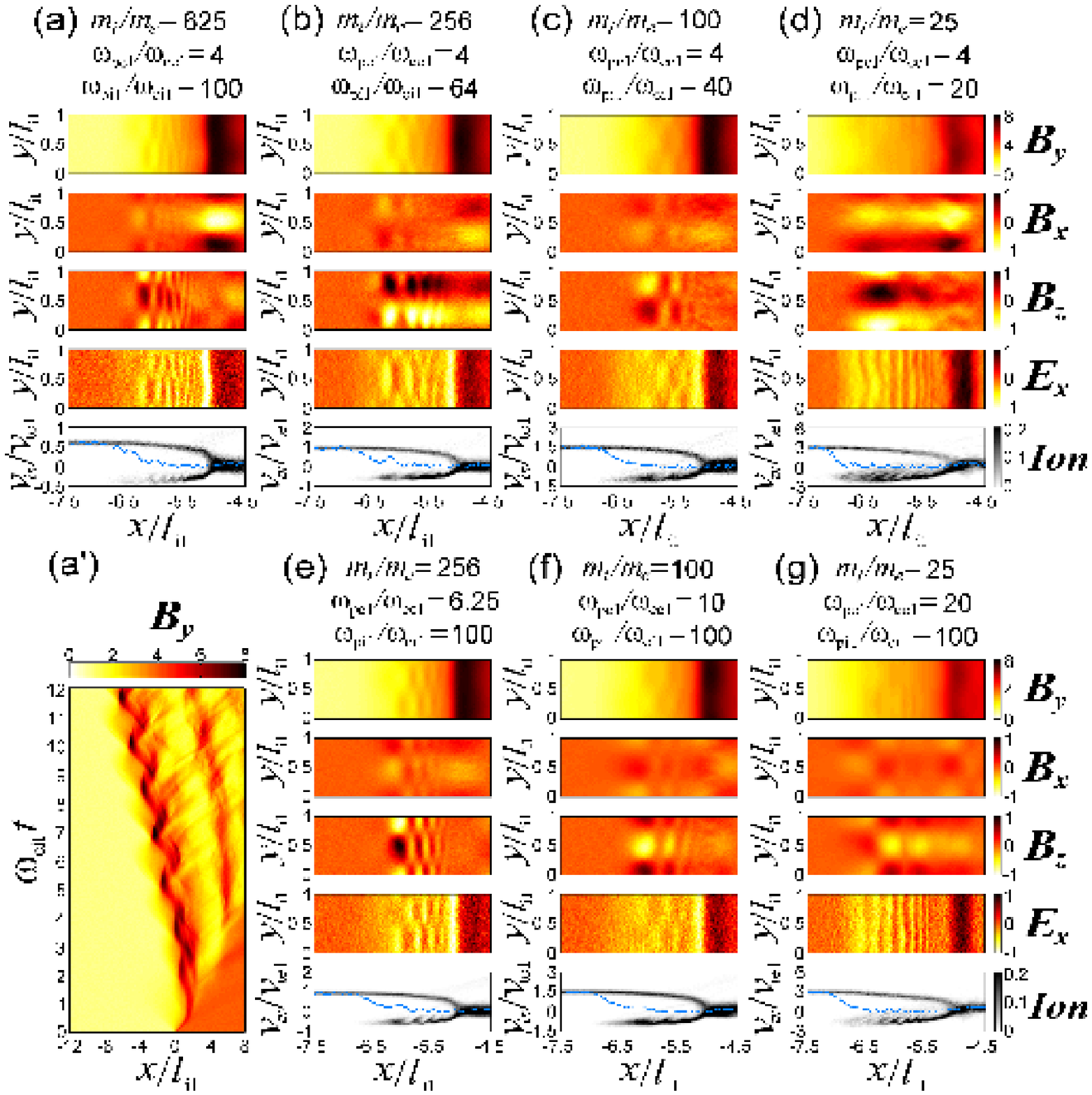}
\caption{
The evolution of the perpendicular shock 
in Runs A-G. 
(a-g) Magnetic field $B_x$, $B_y$, $B_z$ components, 
electric field $E_x$ component, and corresponding 
$x-v_x$ phase-space distribution function of ions 
at $\omega_{ci1}t=$ 10.2 in Runs A-G.  
The solid line in the phase-space plot indicates the electron bulk velocity. 
(a') Tangential magnetic field $B_y$ averaged over $y$ 
as a function of position $x$ and time $t$ 
in Run A. 
The position and time are normalized by 
$l_{i1}$ and $1/\omega_{ci1}$, 
respectively. 
The magnitude of magnetic field is 
normalized by the initial upstream magnetic field $B_{y01}$. 
The magnitude of electric field is 
normalized by the motional electric field $E_{z0}$.  
}
\end{figure*}

%\subsection{Fourier analysis}

In Figures 2 and 3, we show spectra of $B_z$ and $E_x$ components, 
respectively, for Runs A-G. 
These spectra are obtained by Fourier transformation in the range of 
$-6.5 \le x/l_{i1} \le -5.5$, 
$ 0 \le y/l_{i1} \le 1$, and 
$9.8 \le \omega_{ci1}t \le 10.6$. 
The intensity of $B_z$ and $E_x$ is normalized by 
$B_{y01}$ and $E_{z0}$, respectively. 
In the top panels, 
the $\omega-k_x-k_y$ spectrum is 
reduced to the $k_x-k_y$ space by integrating over $\omega$. 
In the bottom panels,
the $\omega-k_x-k_y$ reduced to the $\omega-k_x$ space 
integrating over $k_y$. 

The spectra of both $B_z$ and $E_x$ show that 
there are several wave modes. 
An ion-scale component exists at 
$\omega \sim 0$, $k_x \sim 0$ and $k_y \sim 0$, 
which may correspond to L-mode ion-cyclotron waves 
due to the ion temperature anisotropy.  
However, this wave mode is not focused on 
in the present study because the present Fourier analysis 
cannot resolve ion scales. 

In all the simulation runs, waves with an electron-scale wavelength 
are excited almost in the shock-normal direction ($k_y \sim 0$). 
The excited waves propagate mostly in both ($\pm x$) directions, 
but the waves propagate in the $-x$ direction 
are dominant. 
The phase velocity of the excited waves in the $-x$ direction 
estimated from the $\omega-k_x$ spectra 
corresponds to the drift velocity of the reflected ions. 
The phase velocity of the excited waves in the $+x$ direction  
corresponds to the drift velocity of the incoming ions.

In all the runs, 
the frequency of the $B_z$ component is lower than 
the upstream electron cyclotron frequency $\omega_{ce1}$ 
but is higher than the upstream lower hybrid resonance frequency 
($\omega_{LHR1}/\omega_{ce1} \sim $ 0.04 in Run A, 
 0.06 in Runs B and E, 0.1 in Runs C and F, and 0.2 in Runs D and G). 
These properties agree well with the scenario that 
whistler mode waves propagating in the direction 
oblique to the upstream magnetic field 
are excited by the MTSI due to ion beams 
(incoming or reflected ions) 
across the shock magnetic field. 
In Runs D, F, and G, 
there are also enhancements of the $E_x$ component 
at frequencies higher than the 
upstream electron cyclotron frequency $\omega_{ce1}$ 
but is lower than the upstream upper hybrid resonance frequency 
($\omega_{UHR1}/\omega_{pe1}=4.1$ in Run D, 10.0 in Run F, 
and 20.0 in Run G). 
These properties agree with the scenario that 
electron cyclotron harmonic waves 
propagating in the direction almost perpendicular to 
the upstream magnetic field excited by the ECDI due to ion beams 
(incoming or reflected ions) 
across the shock magnetic field. 
These results suggest that 
the ECDI is generated when the ion-to-electron mass ratio 
is relatively small.

\subsection{Comparison with linear analysis}

\begin{figure*}[p]
\center
\includegraphics[width=1.0\textwidth]{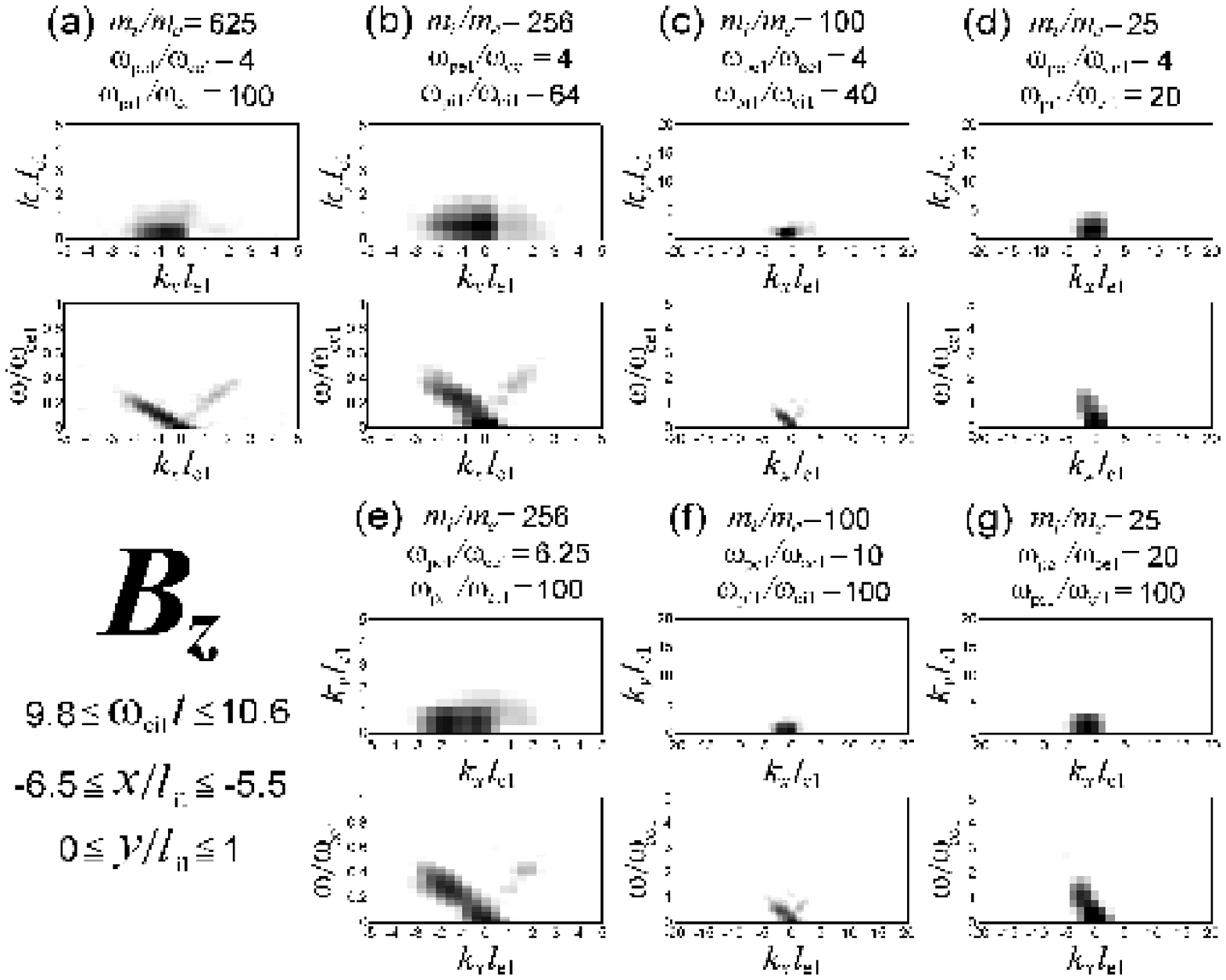}
\caption{
Spectra of $B_z$ component for Runs A-G. 
The magnitude is plotted in a logarithmic scale 
from $10^{-3}B_{y01}$ to $10^{-1}B_{y01}$. 
}
\end{figure*}

\begin{figure*}[p]
\center
\includegraphics[width=1.0\textwidth]{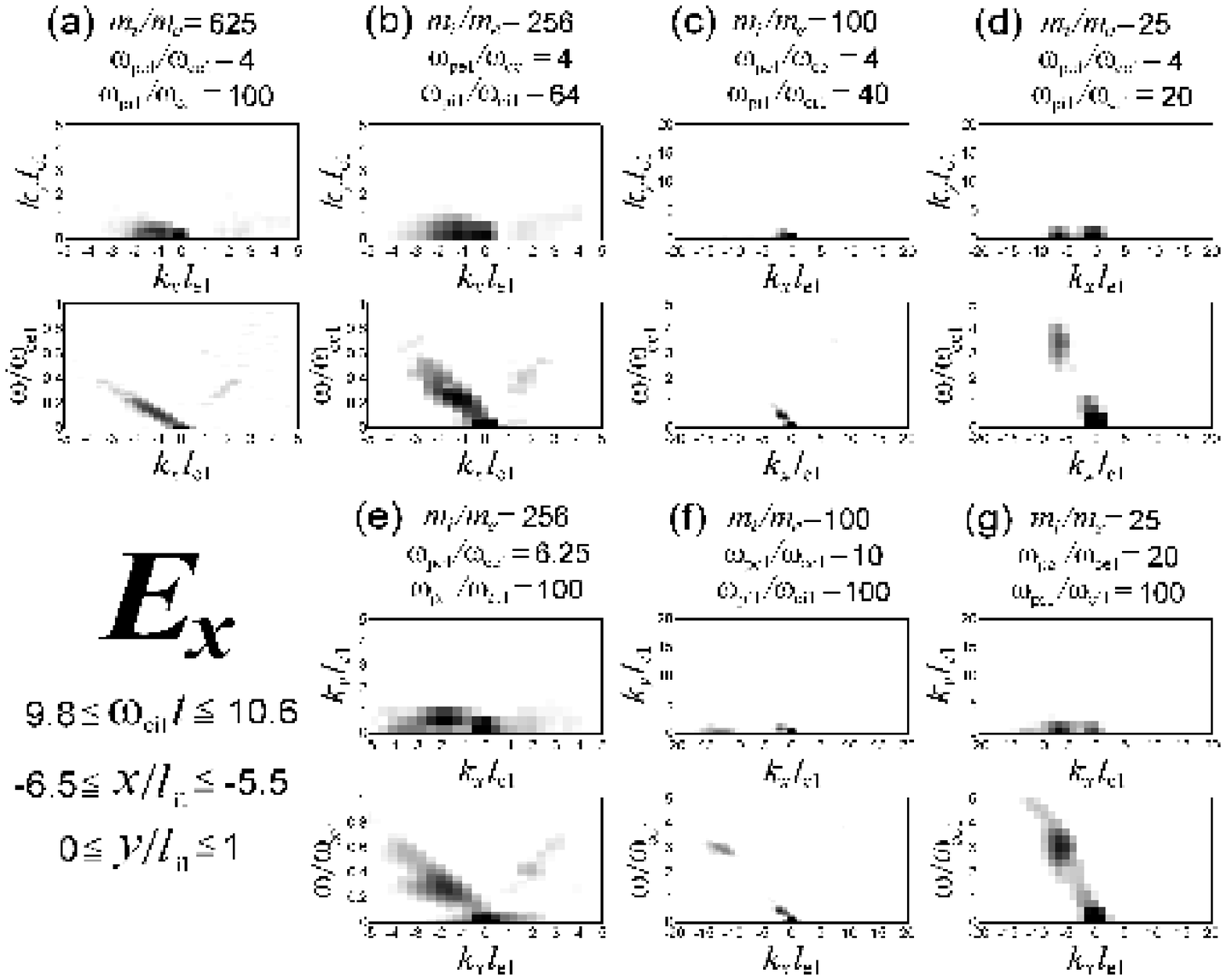}
\caption{
Spectra of $E_x$ component for Runs A-G 
with the same format as Fig.2. 
The magnitude is plotted in a logarithmic scale 
from $10^{-3}E_{z0}$ to $10^{-1}E_{z0}$. 
}
\end{figure*}

A previous linear analysis and numerical simulation 
by Matsukiyo and Scholer \cite{Matsukiyo_2006}  
suggested that there are two-types of instabilities 
due to both incoming and reflected ions. 
Hereafter, we refer to an instability due to incoming ions as 
``instability-1'', 
and refer to an instability due to reflected ions as 
``instability-2''. 

In order to analyze the linear dispersion relation at the shock foot, 
we approximate the velocity distribution function 
with Maxwellian velocity distributions, that is given by
\begin{equation}
f(v) = \frac{n}{\sqrt{2\pi}V_t}\exp\left[-\frac{(v-V_d)^2}{2V_t^2}\right].
\end{equation}
For simplicity, we focus on the velocity distribution function 
only in the shock-normal direction ($x$ component). 
The velocity distribution function $f(v_x)$ is modeled 
by the three-component plasma consisting of 
incoming ions (i1), reflected ions (i2) and bulk electrons (e). 
Figure 4 shows the typical velocity distribution functions of 
ions and electrons at the shock foot 
just before the generation of instabilities in Runs A-G. 
Table 2 shows detailed parameters for 
the Maxwellian velocity distribution functions estimated from Fig.4. 
The velocity, density and frequency 
are normalized by the upstream electron thermal velocity $V_{te1}$, 
the upstream density $n_{01}$, and 
the upstream electron plasma frequency $\omega_{pe1}$, respectively. 

\begin{figure*}[p]
\center
\includegraphics[width=1.0\textwidth]{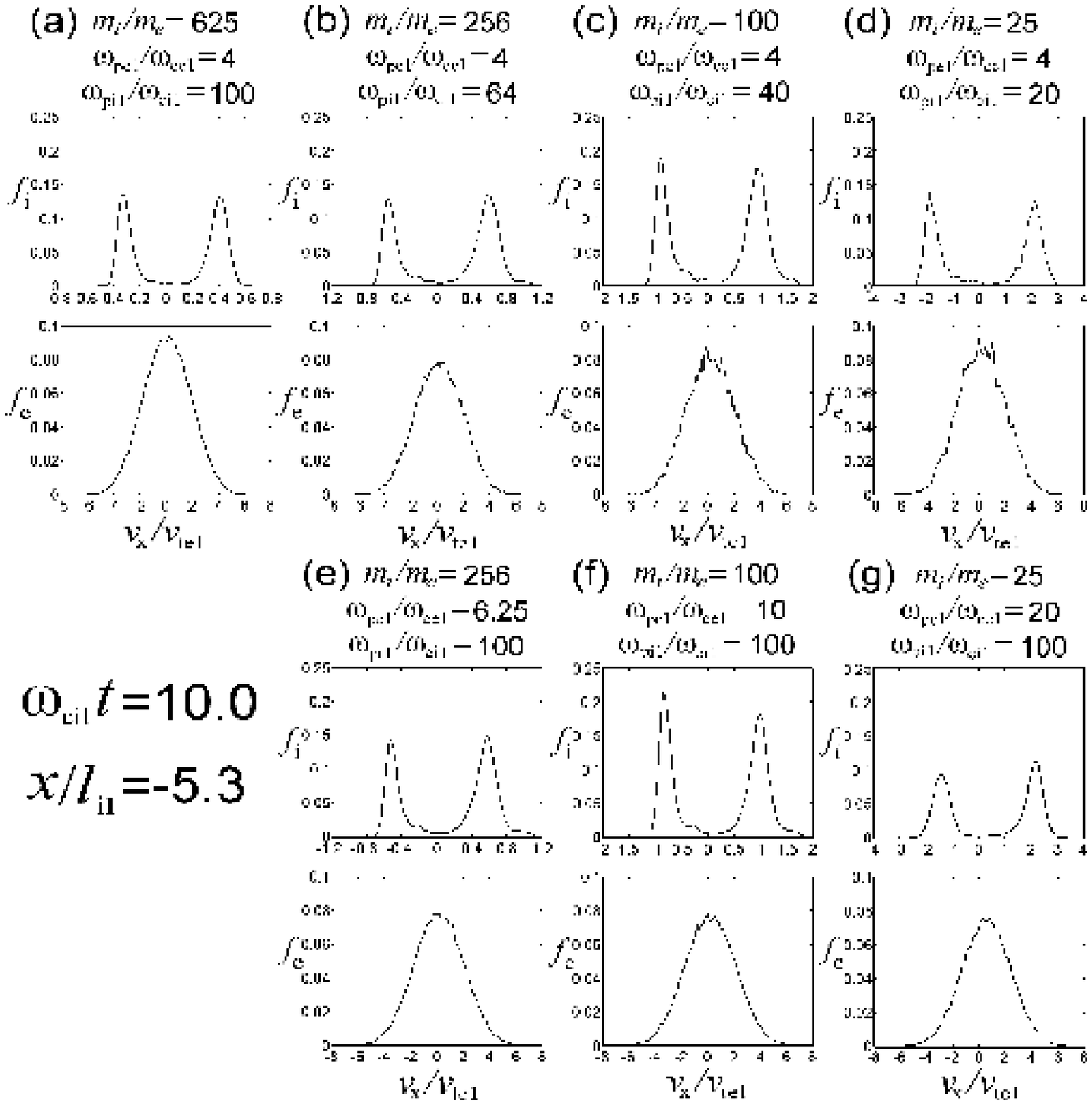}
\caption{
Reduced velocity distribution functions of ions and electrons 
($f_i(v_x)$ and $f_e(v_x)$) averaged over $y$ 
at the shock foot 
just before the generation of instabilities in Runs A-G. 
}
\end{figure*}

\begin{table*}[t]
\caption{
Parameters for the Maxwellian velocity distribution functions 
estimated from Fig.4. 
The velocity, density and frequency 
are normalized by the upstream electron thermal velocity $V_{te1}$, 
the upstream density $n_{01}$, and 
the upstream electron plasma frequency $\omega_{pe1}$, respectively. 
}
\begin{tabular}{c|c|c|c|c|c|c|c|c|c|c|c|c}
Run \ & 
\ \ $V_{di1}$ \ \ & \ \ $V_{ti1}$ \ \ & \ \ $n_{i1}$ \ \ &
\ \ $V_{di2}$ \ \ & \ \ $V_{ti2}$ \ \ & \ \ $n_{i2}$ \ \ &
\ \ $V_{de} $ \ \ & \ \ $V_{te} $ \ \ & \ \ $n_{e} $ \ \ &
\ \ $\omega_{pe}$ \ \ & \ \ $\omega_{ce}$ \ \ \\ \hline
A & 0.42 & 0.06 & 1.68 & -0.32 & 0.05 & 1.38 & 0.08 & 1.94 & 3.06 & 1.75 & 0.92 \\
B & 0.60 & 0.11 & 1.68 & -0.52 & 0.07 & 1.04 & 0.17 & 1.76 & 2.72 & 1.65 & 0.92 \\
C & 0.96 & 0.17 & 1.71 & -0.87 & 0.11 & 1.27 & 0.22 & 1.80 & 2.98 & 1.73 & 0.92 \\
D & 2.12 & 0.28 & 1.70 & -1.85 & 0.21 & 1.41 & 0.32 & 1.75 & 3.11 & 1.76 & 0.92 \\
E & 0.58 & 0.10 & 1.72 & -0.52 & 0.06 & 1.10 & 0.15 & 1.89 & 2.82 & 1.68 & 0.59 \\
F & 0.99 & 0.15 & 1.68 & -0.81 & 0.10 & 1.37 & 0.21 & 1.91 & 3.05 & 1.75 & 0.37 \\
%G & 2.03 & 0.30 & 1.68 & -1.43 & 0.29 & 1.68 & 0.30 & 1.95 & 3.36 & 1.83 & 0.18 \\
G & 2.15 & 0.30 & 1.68 & -1.42 & 0.30 & 1.40 & 0.56 & 1.86 & 3.08 & 1.75 & 0.18 \\
\end{tabular}
\end{table*}

The actual dispersion relation is solved by assuming that 
the ions are unmagnetized and the electrons are magnetized 
as in the previous studies \cite{Matsukiyo_2003,Umeda_2012}. 
It should be noted that although 
both ion and electron components have 
a certain shift in the $v_z$ space (not shown), 
the effect of the drift velocity in the $z$ direction 
is not included in the present analysis.

Figure 5 denotes frequencies (upper panel) and 
growth rates (lower panel) of the instability-1 ($k > 0$) and 
instability-2 ($k < 0$) below the electron cyclotron frequency 
for a variety of wave propagation 
angles with respect to the ambient magnetic field, 
$\theta_{B_k}$, 
as a function of $k_x = k\sin\theta_{B_k}$ and 
$k_y= k\cos\theta_{B_k}$. 
For the above parameters, both of the two instabilities are unstable 
at the phase velocities corresponding to 
the drift velocities of incoming and reflected ions. 
In all the runs, obliquely propagating waves with a frequency $\omega < \omega_{ce1}$
are unstable, which corresponds to 
the MTSI-1 due to incoming ions and the MTSI-2 due to reflected ions. 
We found that the MTSIs have a maximum growth rate at $k_y l_{i1} \sim 4$. 
Although the linear growth rates of both MTSI-1 and MTSI-2 
are of the same order in the linear analysis, 
the MTSI-2 is dominant in all the simulation runs. 
One major reason for this is that the wavenumber resolution 
in the shock tangential ($y$) direction is 
$\Delta k_y = 6.28/l_{i1} \sim 0.25/l_{e1}$ in Run A, 
$\sim 0.39/l_{e1}$ in Runs B and E, 
$\sim 0.63/l_{e1}$ in Runs C and F, and 
$\sim 1.26/l_{e1}$ in Runs D and G, 
so that the simulation system allows the excitation of waves 
only with a larger wavenumber ($k_y =k_x/\tan{\theta_{B_k}} > \Delta k_y$) 
\cite{Umeda_2012}. 
The maximum growth rate occurs at 
$\omega/\omega_{ce1} \sim 0.14$, $k_xl_{e1} \sim 1.26$ in Run A, 
$\omega/\omega_{ce1} \sim 0.26$, $k_xl_{e1} \sim 1.57$ in Runs B and E,  
$\omega/\omega_{ce1} \sim 0.47$, $k_xl_{e1} \sim 1.98$ in Runs C and F, 
which are almost in agreement with the simulation result. 
For Runs D and G, the system does not have 
enough frequency and wavenumber resolutions to 
directly compare the theory and simulation. 
The result suggests that the MTSI-2 is generated 
for a wide range of the mass ratio, 
although the growth rate of the MTSIs becomes smaller 
as the mass ratio becomes smaller.

\begin{figure*}[t]
\center
\includegraphics[width=1.0\textwidth]{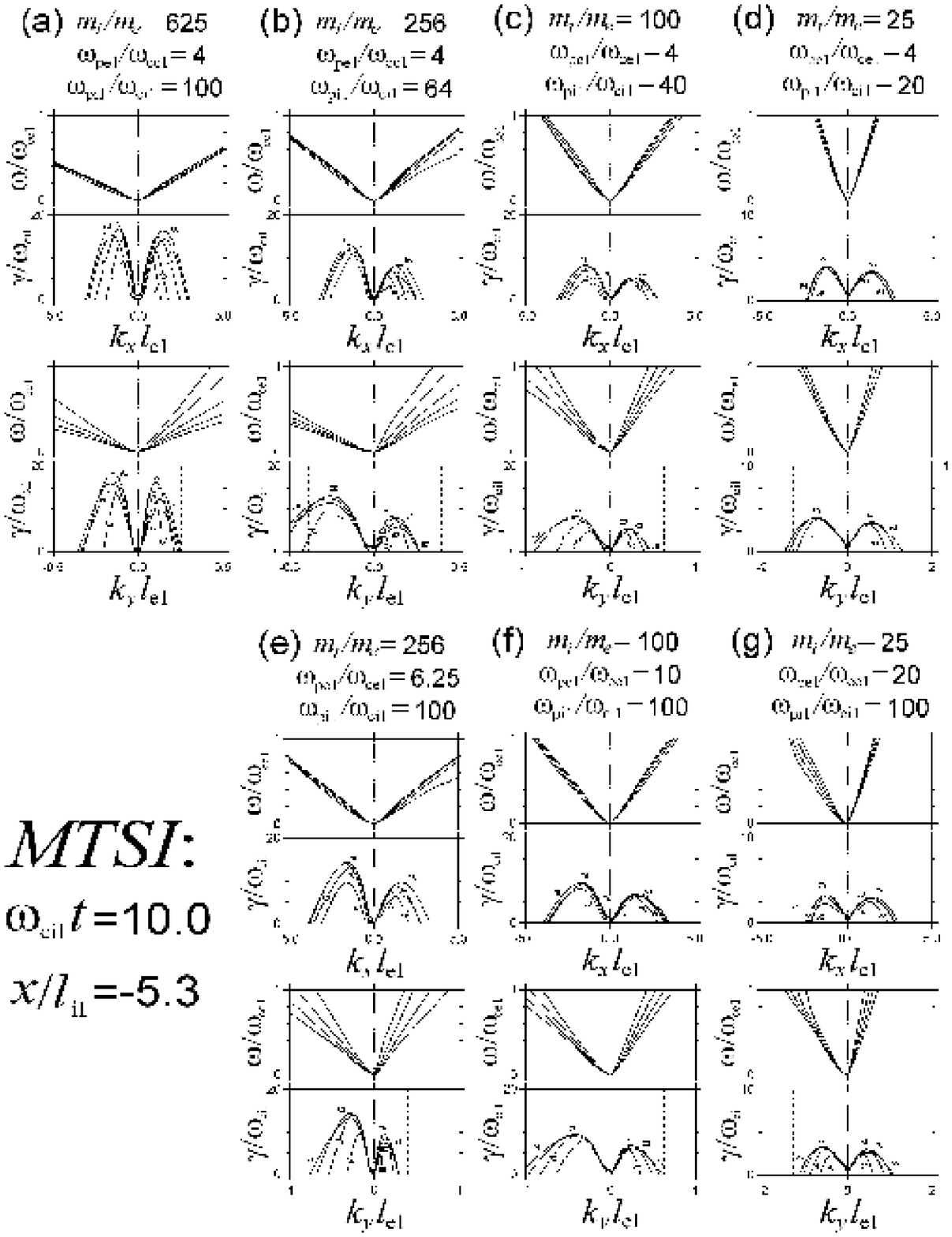}
\caption{
Linear dispersion relations of 
the MTSI-1 ($k_x > 0$) and  MTSI-2 ($k_x < 0$) 
for Runs A-G 
based on the velocity distribution functions in Fig.4. 
Frequencies and growth rates are plotted 
for a variety of wave propagation angles with respect 
to the ambient magnetic field, $\theta_{B_k}$, 
as a function of $k_x = k\sin\theta_{B_k}$ and 
$k_y= k\cos\theta_{B_k}$. 
}
\end{figure*}

% ECDI

Figure 6 denotes frequencies (upper panel) and 
growth rates (lower panel) of the instability-1 ($k_x > 0$) and 
the instability-2 ($k_x < 0$) above the electron cyclotron frequency 
as a function of $k_x$ for Runs C, D, F, and G. 
For the above parameters, both of the two instabilities are unstable 
at the phase velocities corresponding to 
the drift velocities of incoming and reflected ions. 
In theses runs, waves propagating in the direction perpendicular 
to the shock magnetic field are unstable at $\omega \sim n \omega_{ce1}$, 
which corresponds to 
the ECDI-1 due to incoming ions and the ECDI-2 due to reflected ions. 
The maximum growth rate of the ECDI-2 occurs at 
$\omega \sim 3.4 \omega_{ce1}$, $k_xl_{e1} \sim  -6.3$ in Run D, 
$\omega \sim 3.1 \omega_{ce1}$, $k_xl_{e1} \sim -12.5$ in Run F, and
$\omega \sim 3.2 \omega_{ce1}$, $k_xl_{e1} \sim  -8.7$ in Run G, 
which are almost in agreement with the simulation result. 
In Run C, on the other hand, the maximum growth rate of the ECDI-2 
occurs at $\omega \sim 3.1 \omega_{ce1}$, $k_xl_{e1} \sim -10.6$ 
in the linear analysis, 
while there is no wave excitation by ECDI-2 in the simulation. 
This is because Run C does not have enough resolution for 
waves at $k_x = 10/l_{e1} = 1/\lambda_{De1}$. 
We performed a high-resolution 
($\Delta x = 0.1\lambda_{De1}$) 1D simulation, 
and confirmed that the ECDI-2 is weakly generated. 
It should be noted that 
the linear analysis for Runs A, B and E 
shows a positive but small ($\gamma/\omega_{ce1} < 5$) 
growth rate for the ECDIs at a wavelength shorter than 
the electron Debye length ($k_x\lambda_{De1} > 1$). 
However, full particle simulations do not show the generation of the ECDIs 
even with a higher resolution.

Although both ECDI-1 and ECDI-2 have a positive growth rate 
in the linear analysis, 
the ECDI-1 is not generated in these simulation runs. 
Note that the simulation system has enough spatial resolution for ECDI-1 
in Runs D, F and G (see Table 1). 
Here we present several reasons for this 
by using the temporal development of the foot region shown in Fig.7. 
First, we analyze the velocity distribution functions 
near the ramp-side of the foot region, 
while the ECDI-1 is generated from the upstream-side of the foot region. 
Second, waves excited in the ramp-side of the foot region 
quickly reach the shock front 
and the ECDI-1 does not have enough growth time. 
Third, the relative drift between the incoming ions and electrons 
arises later in the broadening phase. 
The incoming ions move through a distance of $2.4l_{i1}$ 
in a time of $0.4/\omega_{ci1}$ (because of $M_A=6$). 
As seen in Fig.7, the structure of the foot region 
quickly changes on the timescale of $0.1/\omega_{ci1}$, 
and the relative drift between the incoming ions and electrons 
exists in a narrow region of $< 2l_{i1}$.

\begin{figure}[ht]
\center
\includegraphics[width=0.5\textwidth]{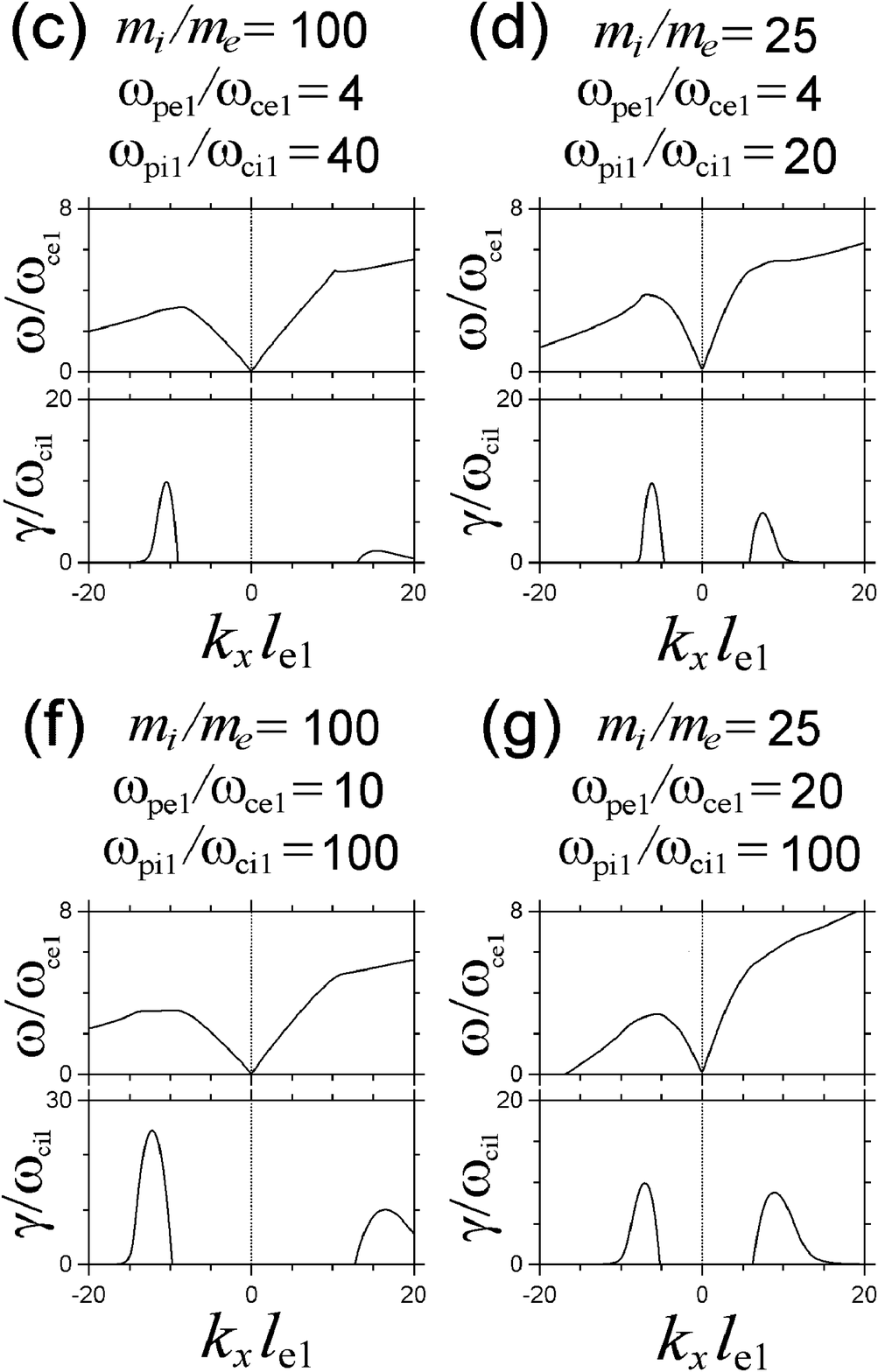}
\caption{
Linear dispersion relations of 
the ECDI-1 ($k_x > 0$) and  ECDI-2 ($k_x < 0$) 
for Runs C, D, F and G 
based on the velocity distribution functions in Fig.4. 
}
\end{figure}
\begin{figure}[ht]
\center
\includegraphics[width=0.5\textwidth]{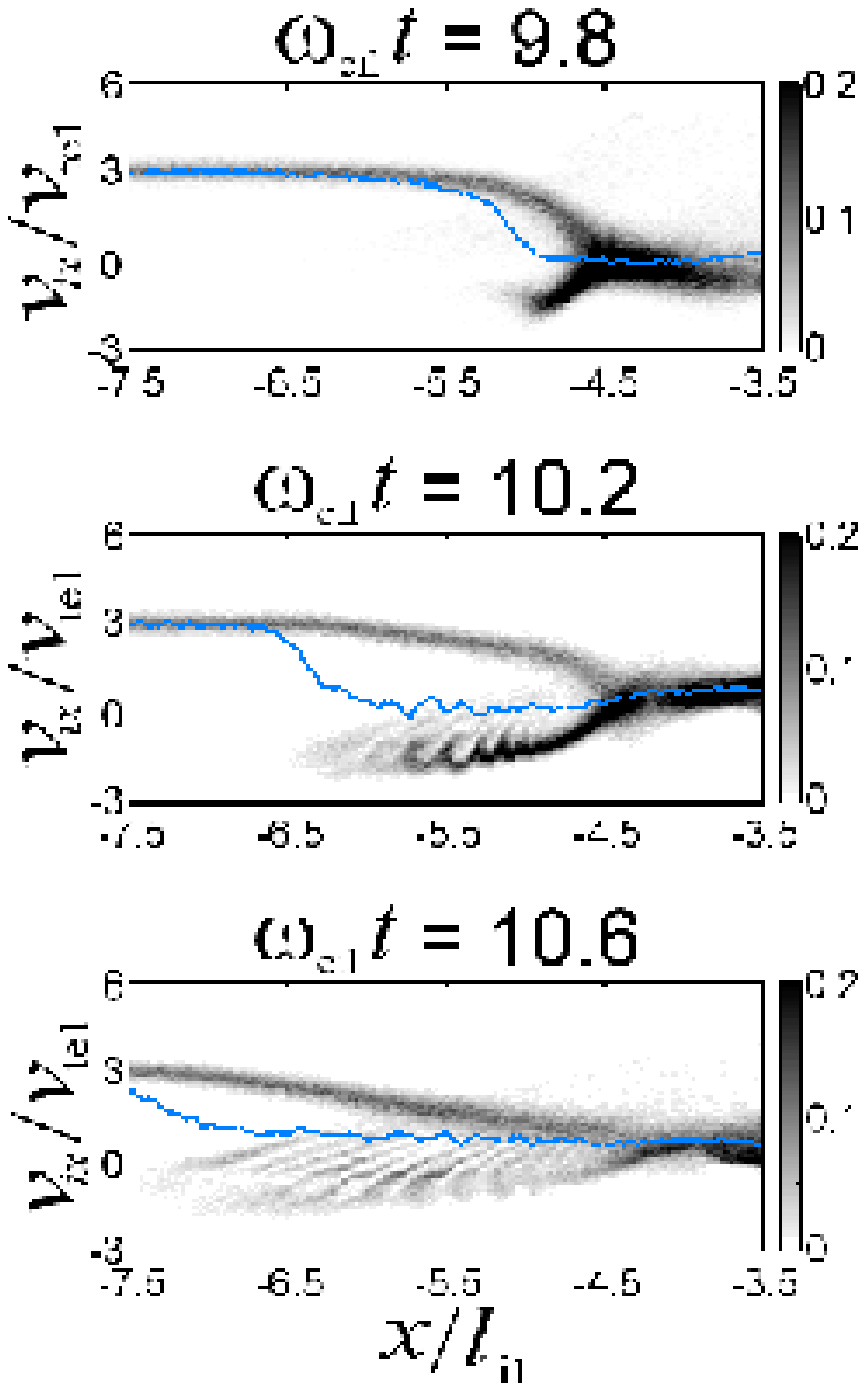}
\caption{
The $x-v_x$ phase-space distribution function of ions 
at different times in Run G. 
The solid line in the phase-space plot indicates the electron bulk velocity. 
}
\end{figure}

\section{Conclusion and Discussion}

The present 2D full particle simulations suggested that 
the ion-to-electron mass ratio affects microinstabilities 
in the foot region of perpendicular collisionless shocks. 
Both simulation result and linear analysis clearly show that 
electromagnetic oblique whistler waves due to the MTSI are dominant 
with a large mass ratio, 
while electrostatic ECH waves due to the ECDI are dominant 
with a small mass ratio.  

The electrostatic ECH waves contribute to the generation of 
nonthermal population of electrons \cite{Umeda_2009}. 
The electromagnetic oblique whistler waves contribute to 
the electron heating in the direction parallel to magnetic fields, 
which slightly modifies the shock jump condition. 
The influence of the excited waves to ion heating 
is small in the present study.

% comparison with Matsukiyo and Scholar

It should be noted that the MTSIs are generated even with small mass 
ratios (Runs B-G), which is seemingly inconsistent with 
the previous expectation \cite{Matsukiyo_2003}. 
However, the previous linear analysis by Matsukiyo and Scholer 
\cite{Matsukiyo_2003} was done by using local (foot) plasma parameters 
as normalization factors, although the linear 
analysis here utilizes the upstream plasma parameters as 
normalization factors. The period of reformation is of the 
order of upstream ion gyro period which is often about 
three times longer than a local ion gyro period estimated 
by using the local magnetic field value in the foot. 

In most of the cases ($m_i/m_e \ge 100$: except for Runs D and G), 
the MTSIs are predominant in terms of wave power compared with the ECDIs. 
As pointed out in the previous studies \cite{Lampe_1972,Wu_1984}, 
ECDI saturates at a low level. 
Furthermore, it is confirmed in a two dimensional periodic (local) simulation 
that the MTSI finally survives even though the ECDI has a larger 
growth rate \cite{Matsukiyo_2006}
if MTSI and ECDI are simultaneously present. 
Here, we confirmed such properties 
in our self-consistent 2D shock simulations.

% future work

In the shock-tangential direction, 
large-amplitude fluctuations on the ion inertial scale 
are commonly exist at the shock front, 
which are known as the ``ripples'' \cite{Winske_1988}. 
Recent full particle simulations with a large system size 
suggested that shock-front ripples change the structure of shocks and 
enhance microinstabilities at the shock front 
\cite{Umeda_2009,Umeda_2010,Umeda_2011,Umeda_2012}. 
In the present study, however, 
the generation of the ripples is not included 
because the system length 
in the shock-tangential direction is too short. 
The present linear analysis shows that 
the MTSIs have the maximum growth rate at $k_{||} l_{i1} \sim 4$, 
which suggests that the oblique whistler waves can couple with 
L-mode ion-cyclotron waves at the shock overshoot 
that are a seed perturbation of the ripples. 
A large-scale 2D simulation including shock-front ripples 
is currently undertaken 
to understand the competition between the ripples and microinstabilities, 
and is reported in a future paper.

\begin{acknowledgments}
This work was supported by MEXT/JSPS under 
Grant-in-Aid for Scientific Research on Innovative Areas No.21200050 
and Grant-in-Aid for Young Scientists (B) 
No.22740323 (S. M.) and No.21740184 (R. Y.). 
The computer simulations were performed on 
the DELL PowerEdge R815 supercomputer system at 
the Solar-Terrestrial Environment Laboratory (STEL) 
and the Fujitsu FX1 and HX600 supercomputer systems 
at the Information Technology Center (ITC), Nagoya University, 
as HPC joint research programs at STEL, ITC, and Joint Usage/Research 
Center for Interdisciplinary Large-scale Information Infrastructures (JHPCN). 
\end{acknowledgments}

\end{document}